\documentstyle[12pt]{article}

\newcommand{\EQ}{\begin{equation}}
\newcommand{\EN}{\end{equation}}
\def\ov{\overline}
\def\al{\alpha}
\def\be{\beta}

\def\la{\lambda}

\def\to{\rightarrow}

\def\mev{\; {\rm MeV}}

\def \prl {Phys. Rev. Lett. }
\def \plb {Phys. Lett. B }
\def \npb {Nucl. Phys. B }
\def \prd {Phys. Rev. D }

\def \nue{\nu_e}

\def \et {E_{\rm th}}

\newcommand{\bea}{\begin{eqnarray}}
\newcommand{\eea}{\end{eqnarray}}
\newcommand{\bean}{\begin{eqnarray*}}
\newcommand{\eean}{\end{eqnarray*}}

\newcommand{\matr}[9]{\left(\begin{array}{ccc}{#1}&{#2}&{#3}\\{#4}&{#5}&{#6}\\
{#7}&{#8}&{#9}\end{array}\right)}
\newcommand{\mcirc}{\stackrel{\circ}{m}}

\begin{document}
\topmargin 0pt
\oddsidemargin=-0.4truecm
\evensidemargin=-0.4truecm
\renewcommand{\thefootnote}{\fnsymbol{footnote}}
\setcounter{page}{1}
\begin{titlepage}
\begin{flushright}
INFN FE-12-95 \\
hep-ph 9507393 \\
July 1995
\end{flushright}
\vspace{0.7cm}
\begin{center}
{\Large \bf Just-So Oscillation:  as Just as MSW? }
\vspace{1.2cm}

{\Large Zurab G. Berezhiani} \footnote{E-mail:
berezhiani@ferrara.infn.it}\\
\vspace{0.2cm}
{\em Istituto Nazionale di Fisica Nucleare, sezione di Ferrara,I-44100
Ferrara,
Italy\\

and\\

Institute of Physics, Georgian Academy of Sciences, Tbilisi
380077, Georgia}\\
\vspace{0.4cm}
and \\
\vspace{0.4cm}
{\Large Anna Rossi} \footnote{E-mail:
rossi@ferrara.infn.it } \\
\vspace{0.2cm}
{\em
Instituto de Fisica Corpuscolar - C.S.I.C.,
Universitat de Valencia \\
46100 Burjassot, Valencia, Spain} \\

\end{center}
\begin{abstract}
The neutrino long wavelength (just-so) oscillation is reconsidered
as a solution to the solar neutrino problem.
In the light of the presently updated results of the four
solar neutrino experiments, the data fit in the just-so scenario
substantially improves and becomes almost as good as in the MSW
scenario. Surprising result of our analysis is that best fit
is achieved when the oscillation occurs only between two neutrino
states: switching on the oscillation into third neutrino increases
the $\chi^2$ value.
Namely, we consider the vacuum oscillation scenario
in the three-neutrino system (4 parameters)
and  find out that the $\chi^2$ minimum is always
achieved in the {\it two} parameter subspace in which actually only
{\it two} neutrino states oscillate.
This holds in the framework of any solar model with relaxed
prediction of the various neutrino fluxes.
The possible theoretical implications of this observation
are also discussed.

\vspace{0.4cm}

\end{abstract}

\end{titlepage}
\renewcommand{\thefootnote}{\arabic{footnote}}
\setcounter{footnote}{0}
\newpage

There are strong arguments to believe that
the Solar Neutrino Problem (SNP), i.e. the deficit of the solar
neutrino fluxes indicated by four operating experiments
\cite{Devis1,Kamioka,Gallex1,Sage1} as compared to the predictions of
the Standard Solar Models (SSM) \cite{BU,BP1}, cannot be
explained by the nuclear/astrophysical reasons.
Namely, the results of different detectors cannot be reconciled
among each other by varying the SSM  parameters as are
the central solar temperature, nuclear cross sections etc. \cite{BB}.
The problem is essentially related to the peculiar energy dependence
needed for the suppression of different solar neutrino components,
so that the intermediate energy $^7Be$ neutrinos are to be killed more
than the lower energy ($pp$) or higher energy ($^8B$) neutrinos.
This points that the SNP is rather due to the "non-standard" neutrino
properties.  The most natural and plausible possibility
is the oscillation of the solar $\nu_e$ into
another type of neutrino $\nu_x$.

The neutrino oscillation picture can provide the necessary energy
dependence in two regimes, which are known as the
Mikheyev-Smirnov-Wolfenstein (MSW) \cite{MSW} and the long wavelength
vacuum oscillation (so called just-so) \cite{GP} scenarios.
The MSW resonant conversion in matter offers a natural possibility
of selective strong reduction of the $^7Be$ neutrinos and thus appears
to be the most attractive and elegant solution to the SNP.
It provides a very good fit of the experimental data,
implying the mass range $\delta m^2\sim 10^{-5}\; {\mbox eV}^2$ and
small mixing angle, $\sin^2 2 \theta\sim 10^{-2}$ \cite{KS}.
The just-so scenario,
with the oscillation wavelength comparable to the sun-earth distance,
needs $\delta m^2$ of about $10^{-10}\,{\mbox eV}^2$
and large mixing angle, $\sin^2 2\theta \sim 1$
\cite{KP,BR},
which parameter range can be naturally generated by non-perturbative
quantum gravitational effects \cite{BEG,ABS}.

Both the MSW and just-so mechanisms are theoretically
motivated\footnote{In the language of the Standard Model (SM)
the neutrino masses emerge through the higher order operators of the
type $\frac{1}{M}(l_iCl_j)HH$, where $l_i$ ($i=1,2,3$) and $H$ are
respectively the lepton and Higgs doublets and $M$ is some regulator scale.
(For example, the famous ``seesaw'' scenario
effectively reduces to these operators with $M$ of
the order or the right-handed neutrino mass).
Then the neutrino mass range needed for the just-so scenario
corresponds to the Planck scale, $M\sim 10^{19}\; {\mbox GeV}$, whereas
the  MSW scenario requires $M$ to be of  the order of the supersymmetric
grand unification scale, $M\sim 10^{16}\; {\mbox GeV}$. }
and so far the experimental data do not allow
to discriminate them. Hence, both these scenarios
should be considered as realistic candidates to the SNP solution.
However, the peculiar "just-so" predictions
as are the seasonal time variation of the neutrino signals and
the specific deformation of the original solar neutrino
spectra \cite{KP,BR}, will allow to test the long wavelength oscillation
scenario at the future real time detectors like
Superkamiokande, SNO, Borexino, etc.,
and to discriminate it from the MSW picture.



In the present paper we present an updated analysis of the just-so
scenario. The earlier analysis \cite{KP,BR} has shown
that the experimental data fit in this scheme
is acceptable, but somewhat worse than that of MSW picture.
However, after publication of
these papers the experimental data have been changed. In particular,
the result of the Homestake experiment has been recalibrated
\cite{Devis1} and the new data of GALLEX experiment became available
\cite{Gallex1}. On the other hand, the
SSM predictions have been reconsidered by taking into
account the helium and heavy element diffusion \cite{BP1}.
The changes are small and rather confirm the stability
of both the experimental and theoretical results. Nevertheless, as we
show below, these small changes make the "just-so" data fit
substantially better so that it becomes
almost as good as in the MSW picture.

Another issue which we address in this paper is the following.
Up to now, in the literature all the analysis of
the vacuum oscillation scenario was performed
within the simplified case of two neutrino ($2\nu$) system,
which case employs only two
parameters: the $\nu_e - \nu_x$ mixing angle $\theta$ and the
mass square difference $\delta m^2$ between two neutrino eigenstates.
However, the nature knows that there exist three neutrino species
and if their masses are originated by the same mechanism, then
the relevant case should be the oscillation picture in the $3\nu$ system,
in which case the solar neutrino data are determined by four
parameters.
Here explore the full parameter space for the 3
neutrino just-so oscillations. As we will see, surprisingly the
introduction of additional parameters dos not improve the data fit and the
best description of the existing data is essentially provided by the
$2\nu$ oscillation, i.e. by the case in which the oscillation into
the third neutrino is decoupled. We briefly discuss the implications
of this observation for the structure of the neutrino mass matrix.

\vspace{3mm}

We perform our analysis in the spirit adopted earlier for the MSW
picture in refs. \cite{KS} and for the just-so scenario
in our previous paper \cite{BR}.
Namely, instead of studying the just-so scenario in the context of
one particular SSM with including in the $\chi^2$ analysis the
theoretical uncertainties, we prefer to scan various realistic SSMs.
We take as a reference model
the recent Bahcall and Pinsonneault (BP) SSM \cite{BP1}
without the underlying theoretical uncertainties,
while the variations of the SSM parameters are taken into account
by parametrizing various neutrino fluxes as $\phi^k = f_k \phi^k_0$
($k=^8B,~ ^7Be,~ ^{13}N,~ ^{15}O,~ pep$ and $pp$),
where the subscript `0' denotes
the values predicted by the BP SSM \cite{BP1} and the factors $f_k$
account for theoretical uncertainties.
Thus, the cases of other SSMs \cite{BU}
will be effectively recovered by relaxing $\phi^k$.
In particular, for a SSM characterized by a given set of $f_k$,
the signals of the radiochemical detectors are expected to be
\begin{eqnarray}\label{signals}
&& R^{SSM}_{Cl}= 7.36f_B + 1.24f_{Be} + 0.22f_{pep} + 0.11f_{N} + 0.37f_{O}
{}~~({\rm SNU}) \nonumber \\
&& R^{SSM}_{Ga}= 69.7f_{pp} +
16.1f_B + 37.7f_{Be} + 3.0f_{pep} + 3.8f_{N} + 6.3f_{O} ~~({\rm SNU})
\end{eqnarray}
as compared to the experimental rates
$R^{exp}_{Cl}= 2.55 \pm 0.25$ SNU \cite{Devis1} and
$R^{exp}_{Ga}= 74 \pm 8$ SNU (for the gallium data we use the
weighted average of the GALLEX result $R_{GALLEX}= 77\pm 8\pm 6$ SNU
\cite{Gallex1} and the SAGE one $R_{SAGE}= 69\pm 11\pm 6$ SNU
\cite{Sage1}).
Thus, the observed signal in each detector can be
given as a ratio of the measured rate to that is expected in a
given SSM, $Z_{a}=R_{a}^{exp}/R_{a}^{SSM}$ ($a=Cl,\,Ga$),
and it is a function of the factors $f_k$.
On the other hand,
the Kamiokande result \cite{Kamioka} implies that the ratio
of the experimental rate to the prediction of SSM with given
$f_B$ is
\EQ\label{kamioka}
Z_K = \frac{R^{exp}_K}{R^{SSM}_K} = \frac{1}{f_B}\, (0.45\pm 0.08)
\EN
where obviously the case $f_B=1$ corresponds to the BP SSM \cite{BP1}.

Several numerical studies
have shown that in practice the effects of independent variations
of the metal fraction, the opacities, the astrophysical factor $S_{pp}$
and the age of the sun are well reproduced by  variations of
the central solar temperature $T_c$ according to the
power lows \cite{T-c}:
\begin{equation}\label{power}
\phi^k = \phi^k_{0} \left(\frac{T_c}{T_c^{0}}\right)^{\beta_k}
\end{equation}
Thus, we consider $f_B=(T_c/T_c^0)^{\beta_B}$ as a free parameter
while for other coefficients we correspondingly obtain
$f_k= f_B^{\xi_k}$, where $\xi_k=\beta_k/\beta_B$.
Following ref. \cite{T-c}, we take
$\xi_{pp} = - 0.034 \pm 0.006$, $\xi_{Be} = 0.47 \pm 0.08$,
$\xi_{N} = 1.0 \pm 0.4$, $\xi_{O} = 1.2 \pm 0.4$ and
$\xi_{pep} =0 \pm 0.1$.\footnote{In the following computations we take
the central values neglecting the theoretical uncertainties.
These uncertainties are indeed small for the $pp$ and $Be$ neutrinos.
For the CNO components these are bigger but the contribution of the
latter to the signals  (\ref{signals}) are not substantial. }
In this way, the case of other SSM's will be
effectively recovered by varying $f_B$. For example, by taking
$f_B\simeq 0.7$ the Turk-Chieze and Lopez SSM \cite{BU}
is reproduced.
The factor $f_B$ is varied in
the range $0.4-1.6$ (the lower limit $f_B\simeq 0.4$ is in fact set
by the Kamiokande result), which in fact corresponds to the variation
of the central solar temperature within $(-4 \div +2)~ \%$.

\vspace{3mm}
Let us first consider the vacuum oscillation solution
 in the case of two neutrino flavours:
$\nu_e\to \nu_x$ where $\nu_x$ is $\nu_\mu$ or $\nu_\tau$
(alternatively, it can be also a sterile state $\nu_s$). In this case
the survival probability for solar $\nu_e$'s with energy $E$ depends
on two parameters, the mixing angle $\theta$ and
the mass square difference $\delta m^2$:
\begin{equation}
P(E)= 1-\sin^2 2 \theta \sin^2
\left(\frac{\delta m^2 L_t}{4E} \right)
\label{surp}
\end{equation}
The sun-earth distance depends on time as
$L_t= \ov{L}[1 -\varepsilon\cos (2\pi t/T)]$, where
$\ov{L}=1.5 \cdot 10^{11}$m, $T=1$ yr, and
$\varepsilon=0.0167$ is the ellipticity of the orbit.
Then the time averaged detection rates expected in the radiochemical
experiments read as
\EQ
\label{rate1}
R= \int dE \sigma (E) \sum_k \langle P(E) \phi^k \rangle _{T}\la_k(E)
\EN
where $\sigma(E)$ is the detection cross section, $\phi^k=f_k\phi^k_0$
are the fluxes of the relevant components of the solar
neutrinos ($k=B, Be,$ etc.), $\la_k(E)$ are their
energy spectra normalized to 1,  and $\langle \dots \rangle_T$ stands
for the average over the whole time period $T$.
%
For the Kamiokande detector we have
\EQ
R_K=  \int_{\et} dE \la_B(E) \left[ \langle P(E)  \phi^B\rangle _{T}
 \sigma_{\nue}(E) +\left( \langle \phi^B \rangle _{T}
- \langle P(E)  \phi^B\rangle _{T}\right)
\sigma_{\nu_x}(E) \right]
\label{rate2}
\EN
where $\sigma_{\nu_y}$ $(y=e,x)$ is the $\nu_y e^-$ scattering cross
section (in the case of conversion into sterile neutrino
$\sigma_{\nu_x}=0$) and
$\et= \frac{1}{2}( T_e+ \sqrt{T_e(T_e +2m_e)})$, where $T_e=7.5\mev$
is the recoil electron kinetic energy threshold.

We accept the hypothesis that the solar
neutrino luminosities are constant in time, and
use the averaged data of the chlorine, gallium and Kamiokande
experiments to perform the standard $\chi^2$ analysis for various cases.
We find it instructive to separate the experimental and theoretical
(SSM) uncertainties, and do not include the latter in $\chi^2$ analysis
once these are simulated by varying the factors $f_{k}$.
We define
\EQ\label{chi2}
\chi^2 = \sum_a \left(\frac{R^{exp}_a- R_a}{\sigma_a}
\right)^2
\EN
where the index $a=Cl,Ga,K$ labels three different data
and $\sigma_a$ are the corresponding experimental errors.
Once we describe the three experimental results by means of
two paramemters ($\delta m^2$ and $\sin^2 2\theta$), the number
of the degrees of freedom is 1.

We have performed the $\chi^2$ analysis for various values of $f_B$,
using for the numerical minimization procedure
the MINUIT package provided by  the CERN program library.
The corresponding best fit points of $\chi^2_{min}$
and `1$\sigma$' areas containing the true parameter values with
68 \% probability, once the solution is assumed, are shown in Fig. 1.
We see that compared to the previous analysis \cite{BR}, the fits have
become much better. E.g. in the case $f_{B}=1$ we have
$\chi^2_{min}=1.6$ (versus $\chi^2_{min}=4.4$ obtained in our
previous paper \cite{BR}),
so that the just-so oscillation is allowed as a SNP solution at the
20 \% confidence level (C.L.).
\footnote{ Since the theoretical uncertainties are not included,
here and in the following the quantitative statements on the C.L.
are somewhat informal, and reflect only the ideal situation with
no uncertainties in SSM predicted values of the neutrino fluxes,
exactly fixed by the choice of the factors $f_k$.
For the case of $\nu_x$ being a sterile neutrino we obtain
$\chi^2_{min}=5.5$. }

By varying $f_B$, the relevant range of $\delta m^2$ remains rather
stable, while $\sin^2 2\theta$ tends to be smaller with decreasing
$f_B$. The lowering (increasing) of $f_B$ implies the weakening
(strengthening) of the oscillation. However, as a general tendency,
by decreasing $f_B$ the fit becomes worse: e.g. for $f_B=0.4$
we have $\chi^2_{min}=3.2$, while it improves for $f_B>1$:
e.g. for $f_B=1.6$ we have $\chi^2_{min}=0.9$, which is acceptable
at 34 \% C.L.

Thus our conclusion is that for $f_B\geq 0.7$
the relevant parameter range is limited by the values
$\delta m^2=(0.5-1) \times 10^{-10}~ \mbox{eV}^2$ and
$\sin^2 2\theta = 0.7-1$. For $f_B\leq 0.7$ the parameter
range is somewhat wider (see Fig. 1) but as we have seen in this case the
quality of the fit is less good). The dependence on
$\sin^2 2\theta$ is simpler -- the just so scenario preferes
the large (close to the maximal) mixing,\footnote{According to
ref. \cite{BSS}, such a large mixing puts the just-so oscillation
in contradiction with the observed pattern of the neutrino signal
from SN 1987A. See, however, the recent paper \cite{KK}. } whereas
the
dependence on $\delta m^2$ is more sophisticated, due to the different
oscillation length of solar neutrinos with different energies.
To give some more numerical insight on this dependence,
in the Figs. 2a,b we show $\chi^2$ as a function of $\delta m^2$
for the cases $\sin^2 2\theta=1$ and $8/9$, which correspond
to the cases of the maximally symmetric (democratic) mixing
in the 2 and 3 neutrino system (see below).

\vspace{3mm}

Let us consider now the case of three neutrino mixing, when the neutrino
mass eigenstates $\nu_i$ with masses $m_i$ ($i=,1,2,3$) are related
to the weak eigenstates $\nu_{\al}$ ($\al=e,\mu,\tau$) as
$\nu_{\al}= U_{\al i}\nu_i$, where $U$ is the 3x3 unitary matrix
of the neutrino mixing.
The general expression for the $\nu_\al \rightarrow\nu_\be$
transition probability is:
\EQ\label{3nu}
P_{\alpha\beta}(E) = \left \arrowvert\sum_i U_{\be i} U^{*}_{\al i}
\exp (-{\rm i} E_i t)
\right \arrowvert ^2
\EN


The radiochemical detectors are sensitive only to $\nu_e$ state.
On the other hand,
the Kamiokande experiment does not distinguish between
the contributions of $\nu_\mu$ and $\nu_\tau$, provided that the latter
have no non-standard interactions with electrons
($\sigma_{\nu_\mu}=\sigma_{\nu_\tau}$ in eq. (\ref{rate2})). Hence,
the analysis of the solar neutrino data in the 3$\nu$ case needs only
the expression of the $\nu_e\rightarrow \nu_e$ survival probability:
\EQ\label{surv}
P(E)= 1 -4\left |U_{e1}  U_{e2}\right| ^2 \sin^2
(\frac{\delta_{21} L_t}{4 E})
-4\left| U_{e1} U_{e3}\right | ^2 \sin^2
(\frac{\delta_{31} L_t }{4 E})
-4\left| U_{e2} U_{e3}\right | ^2 \sin^2
(\frac{\delta_{32} L_t}{4 E})
\EN
where $\delta_{ij} = m^2_{i}-m^2_{j}$.
This expression depends on four independent variables which
we choose as $U_{e 2}, U_{e3}$
and $\delta_{21}, \delta_{31}$, whereas
$\delta_{32} = \delta_{31}-\delta_{21}$ and the unitarity constraints
fix $|U_{e1}|^2= 1 - |U_{e2}|^2- |U_{e3}|^2$.
One can use e.g. Maiani-like parametrization by identifying
$U_{e1}=\cos\theta\cos\phi$, $U_{e2}=\sin\theta\cos\phi$ and
$U_{e3}=\sin\phi$.

The $2\nu$ oscillation is a limiting case which is achieved if
the third neutrino is decoupled. E.g. in the limit $U_{e3}=0$
($\sin\phi=0$),
in which case the dependence $\delta_{31}$ is also lost,
we have $4|U_{e1}U_{e2}|^2=\sin^2 2\theta$ and thus
eq. (\ref{surv}) reduces to the familiar expression (\ref{surp}).
On the other hand, when e.g. $\delta_{32}=0$, we can identify
$\delta_{31}=\delta_{21}=\delta m^2$ and still
$4|U_{e1}|^2 (1-|U_{e1}|^2)=\sin^2 2\theta$.

In principle, the presence of four parameters versus only three
experimental data makes the system overdetermined and thereby
the standard meaning of $\chi^2$ analysis is not directly applicable.
However, even in this case the canonical $\chi^2$ function (\ref{chi2})
provides a numerical insight on the deviation of the theoretical
estimates from the experimental data, within their  errors.
Its minimization with respect to all the
four parameters can allow to know whether the $3\nu$ just-so scenario
shows a better ``fit'' (i.e. smaller $\chi^2$)
in some relevant parameter region as
compared to the $2\nu$ case. As smaller the value of $\chi^2$ is,
as closer is the theoretical "just-so" prediction with
experimental data.

The result of our analysis is rather surprising.
Performing numerically the minimization procedure
in the four parameter space, we always find that the best fit point
(for which $\chi^2$ achieves the minimum) is always located
in the two parameter subspace which effectively corresponds
to the $2\nu$ case, i.e. when e.g. $U_{e3}=0$ or $\delta_{32}=0$,
for any feasible value of $f_B$.
The values of the minimal  $\chi^2_{min}$ coincides with that
is obtained above in the 2$\nu$ scheme.
For example, in the case $f_B=1$ we obtain
$\chi^2_{min}=1.6$, which occurs for the parameter values
$|U_{e3}|^2 =0$,
$|U_{e2}|^2 =0.40$ (i.e. $\sin^2 2\theta=0.96$)
and $\delta_{21}= 0.6\cdot 10^{-10}\,$eV$^2$. Of course, there are also
trivially equivalent minima obtained by permutations
$U_{ei}\to U_{ej}$, or by identyfying $\delta_{ij}=\delta_{ik}$.

The feature seems rather obvious
in the limit when say $\delta_{31}$ is taken to be large, i.e.
$m_3\gg 10^{-5}\,$eV.
In this case the two last terms in (\ref{surv}) are in the regime of the
averaged vacuum oscillation, and the $\nu_e$ survival
probablity reads
\begin{equation}\label{averaged}
P(E)= 1 - \frac{1}{2}\sin^2 2\phi -
\cos^4 \phi \sin^2 2\theta \sin^2(\frac{\delta_{21} L_t}{4E})
\end{equation}
Then for non-zero $\phi$ the data fit has to
become worse,
since the contribution of the energy dependent term becomes
smaller which makes it more difficult to reconcile the
Homestake and Kamiokande data. This feature, however,
is not so obvious when $\delta_{31}$ is also in the just-so
($\sim 10^{-10}\,\mbox{eV}^2$) range.


\vspace{3mm}

As we have seen, the best just-so description of solar neutrino data
is achieved when in fact only 2 neutrino states participate the oscillation.
Let us discuss briefly how this situation could be obtained in the
context of various neutrino mass generation schemes.

The neutrino mass range needed for
the just-so oscillation can naturally emerge from the Planck scale
effects \cite{BEG,ABS}.
In the minimal standard model (or SU(5) GUT) the neutrinos are massless
as far as renormalizable interactions are concerned. In other words,
in the absence of the right-handed (RH) neutrino states the lepton
number conservation arises as an accidental global symmetry of the
theory due to the joint requirement of gauge invariance and
renormalizability. However, the global symmetries need not be
respected by nonperturbative quantum gravity effects.
Thus, if the SM (or SU(5)) is a true theory up to the Planck scale,
the neutrino masses can be induced only by the non-renormalizable
operators cutoff by the Planck scale $M_{Pl}$:
\begin{equation}\label{MPl}
\frac{\alpha_{ij}}{M_{Pl}}(l_i H) C(l_j H)
\end{equation}
where $l_i=(\nu_i,e_i)^T$ ($i=1,2,3$)
are the left-handed (LH) lepton doublets,
$H$ is the standard Higgs doublet, and $\alpha_{ij}$ are the order 1
constants. (In fact, this is equivalent to the seesaw mechanism
with RH neutrino states having $\sim M_{Pl}$
Majorana masses and $\sim 1$ Yukawa constants;
in particular, such a situation is automatically obtained in
a supersymmetric grand unified model $SU(6)$ \cite{SU6}).
As a result, the neutrinos get Majorana masses of the order of
$\mcirc~ = v^2/M_{Pl} = 3\cdot 10^{-6}$ eV,
where $v=174\,$GeV is a VEV of the Higgs doublet.
The parameter $\delta m^2 \sim \mcirc^2$ thus naturally emerges
in the range needed for the just-so oscillation.

In general case, this operator induces mass terms for all
three neutrinos $\nu_e, \nu_\mu, \nu_\tau$.
If the matrix $\alpha_{ij}$ of the effective coupling constants is a
general one, then we have the $3-\nu$ oscillation case.
However, it could occur that $\alpha_{ij}$ is a rank-1 matrix,
\begin{equation}\label{P}
\alpha = U^T P U ,~~~~~~~~~~~~~~~~~~P = {\mbox diag}(0,0,1)
\end{equation}
where $U$ is some unitary matrix.
Then only one neutrino eigenstate gets mass of the order of
$\mcirc$, while others stay massless. In this case the neutrino
oscillation picture effectively reduces to the $2\nu$ case.
For example, it was argued in \cite{ABS} that once the operator
(\ref{MPl}) is induced by nonperturbative gravitational effects
(virtual black holes or wormholes), the flavour blindness of
gravity could imply the following pattern for the neutrino mass
matrix
\begin{equation}\label{blind}
\hat{m}^0_\nu=\alpha \mcirc \matr{1}{1}{1} {1}{1}{1} {1}{1}{1} =
U^T \matr{0}{0}{0} {0}{0}{0} {0}{0}{3\alpha \mcirc} U
\end{equation}
in which case we effectively have $2\nu$ oscillation with
$\delta m^2 = 9\alpha^2 \mcirc^2$ and $\sin^2 2\theta = 8/9$,
in good agreement with the data fit in the just-so picture
(see Fig. 2b).

It is however also possible that one of the neutrino eigenstates
acquires a mass larger than $\mcirc$ from  sources different
from the Planck scale induced operators (\ref{MPl}). In particlular,
the neutrino mass matrix may have a structure
\begin{equation}
\hat{m}_\nu= m \hat{T} + \hat{m}_\nu^0 ,~~~~~~ m\gg \mcirc
\end{equation}
where $T$ is some rank-1 matrix. If $T=P$, where $P$ is defined
in eq. (\ref{P}), then the $\nu_\tau$ state acquires mass $m$
and for $m\sim 5-7$ eV it can play a role of the cosmological
hot dark matter component \cite{HDM}.
In this case for the SNP solution remains
the just-so oscillation $\nu_e\to \nu_\mu$ with maximal mixing
$\sin^2 2\theta=1$, which still provides a good fit to
the solar neutrino data (see Fig. 2a).

Alternatively, one can imagine that the matrix $T$ has a democratic structure
 in the space of the $\nu_\mu$ and $\nu_\tau$
weak eigenstates:
\begin{equation}
T=\matr{0}{0}{0} {0}{1}{1} {0}{1}{1}
\end{equation}
In this case the $\nu_\mu\to \nu_\tau$ oscillation emerges
with $\sin^2 2\theta_{\mu\tau}=1$ and $\delta m^2_{\mu\tau}=4m^2$,
which for $m\simeq 5\cdot 10^{-3}$ eV could explain the
the atmospheric neutrino deficit \cite{ANP}, while the
just-so oscillation is still effective in explaining the SNP.

\vspace{3mm}
Concluding, we have seen that the just-so oscillation scenario
provides a very good fit to the recently updated
experimental data on the solar neutrinos, and its present status
is as reasonable as that of the widely popular MSW scenario.
We have shown that best data fit is achieved within the two
neutrino system, and with incorporating the third neutrino the
fit can only get worse. This points that the third neutrino
state can be involved into the games of solving other present
neutrino puzzles as are the atmospheric neutrino problem or
the problem of the cosmological hot dark matter.

\vspace{2cm}


{\bf Acknowledgements.}

\vspace{3mm}

We are grateful to Gianni Fiorentini for illuminating  conversations.
Useful discussions with Giovanni Di Domenico, Barbara Ricci,
Alexei Smirnov and Jose Valle
are also gratefully acknowledged.

\newpage
{\Large \bf Figure Captions }
\vspace{10mm}

Fig. 1.
The long-dashed, dott-dashed, solid, dashed and dotted contours mark
the 68 \% C.L. regions respectively for $f_B$=0.4 , 0.7, 1, 1.3, 1.6,
The corresponding best fit points are shown by diamonds,
with $\chi^2_{min}$ being respectively 3.2, 2.5, 1.6, 1.5, 0.9.

Fig. 2. $\chi^2$ as a function of $\delta m^2$ for a fixed mixing
angle: $\sin^2 2\theta=1$ (a) and $\sin^2 2\theta=8/9$ (b).
The dot-dashed, solid and dashed curves correspond to the
cases $f_B =$ 0.7, 1 and 1.3, respectively.


Fig. 3. $\chi^2$ minimized with respect to parameters $\sin^2 2\theta$
and $\delta_{21}$, as a function of the `13' mixing
$\sin^2 2\phi$, for the case of large $\delta_{31}$
($m_3 \gg \mcirc$). The dot-dashed, solid and dashed curves
correspond to the cases $f_B =$ 0.7, 1 and 1.3, respectively.


\begin{thebibliography}{99}


\bibitem{Devis1} B.T. Cleveland et al.,
Nucl. Phys. B (Proc. Suppl.) {\bf 38} (1995) 47.

\bibitem{Kamioka} Kamiokande Collaboration,
Nucl. Phys. B (Proc. Suppl.) {\bf 38} (1995) 55.

\bibitem{Gallex1} GALLEX Collaboration, P. Anselmann {\em et al.},
preprint LNGS 95/37.

\bibitem{Sage1} SAGE Collaboration, J.S. Nico et al., Proc. 27th
Int. Conf. on High Energy Physics, Glasgow, UK (July 1994).

\bibitem{BU} J. N. Bahcall and R. K. Ulrich,
Rev. Mod. Phys. {\bf 60} (1989) 297;
J.N. Bahcall and M.H. Pinsonneault, Rev. Mod. Phys. {\bf 64} (1992) 885;
S. Turk-Chi$\grave{e}$ze and I. Lopez,
Ap. J. {\bf 408} (1993) 347; S. Turk-Chi$\grave{e}$ze {\it et al.},
Phys. Rep. {\bf 230} (1993) 57;
V. Castellani, S. Degl'Innocenti and G. Fiorentini, Astron. Astrophys.
{\bf 271} (1993) 601.

\bibitem{BP1} J.N. Bahcall and M.H. Pinsonneault, preprint IASSNS-AST 95/24.

\bibitem{BB}
V. Castellani, S. Degl'Innocenti and G. Fiorentini,
\plb {\bf 303} (1993) 68;
V. Castellani {\it et al.}, \plb {\bf 324} (1994) 425;
J.N. Bahcall and H.A. Bethe, \prl {\bf 65} (1993) 2233;
J.N. Bahcall {\it et al.}, preprint IASSNS-AST 94/13 (1994);
J.N. Bahcall, preprints IASSNS-AST 94/14,  IASSNS-AST 94/37 (1994);
S. Bludman {\it et al.}, Phys. Rev. D {\bf 47} (1993) 2220;
Phys. Rev. D {\bf 49} (1994) 3622;
V.S. Berezinsky, Comments Nucl. Part. Phys. {\bf 21} (1994) 249;
A.Yu. Smirnov, preprint DOE/ER/40561-136-INT94-13-01 (1994).


\bibitem{MSW}
S.P. Mikheyev and A.Yu. Smir\-nov,
Yad. Fiz. {\bf 42} (1985) 1441;
Nuovo Cimento {\bf 9C} (1986) 17;
L. Wolfenstein, Phys. Rev. D {\bf 17} (1978) 2369; D {\bf 20} (1979) 2634.

\bibitem{GP}
V. Gribov and B. Pontecorvo, Phys. Lett. {\bf 28} (1967) 493;
J.N. Bahcall and S. Frautschi, Phys. Lett. {\bf 29B} (1969) 623;
 V. Barger, R.J.N. Phillips and K. Whisnant, \prd
{\bf 24} (1981) 538; S.L. Glashow and L.M. Krauss, \plb {\bf 190}
(1987) 199.




\bibitem{BEG} R. Barbieri, J. Ellis and M.K. Gaillard, \plb {\bf 90}
(1980) 249.

\bibitem{ABS} E. Akhmedov, Z. Berezhiani and G. Senjanovi\'c,
\prl {\bf 69} (1992) 3013.

\bibitem{KS} P.I. Krastev and A.Yu. Smirnov, \plb {\bf 338} (1994) 282;
V. Berezinsky, G. Fiorentini and M. Lissia, \plb {\bf 341} (1994) 38;
N. Hata and P. Langacker, \prd {\bf 50} (1994) 632.

\bibitem{KP} P. Krastev and S. Petcov,
\prl {\bf 72} (1994) 1960; preprint SISSA 41/94/EP (1994).

\bibitem{BR} Z.G. Berezhiani and A. Rossi, \prd {\bf 51} (1995) 5229.

\bibitem{T-c} V. Castellani et al., \prd {\bf 50} (1994) 4749.

\bibitem{ABL} A. Acker, A.B. Balantekin and F. Loreti, \prd {\bf 49}
(1994) 328.

\bibitem{BSS} J.N. Bahcall, A.Yu. Smirnov and D.N. Spergel,
\prd {\bf 49} (1994) 1389.

\bibitem{KK} P.J. Kernan and L.M. Krauss, \npb {\bf 437} (1995) 243.





\bibitem{SU6} Z.G. Berezhiani, hep-ph/9503366, \plb (in press).

\bibitem{ANP} Kamiokande Collaboration, Y. Fukuda {\em et al.,}
\plb {\bf 335} (1994) 237.

\bibitem{HDM} R. Shaefer and Q. Shafi, Nature {\bf 359} (1992) 199.



\end{thebibliography}
\end{document}